# Operation of Multi-Pixel Radio-Frequency Superconducting Nanowire Single-Photon Detector Arrays

Steffen Doerner, Student Member, IEEE, Artem Kuzmin, Stefan Wuensch, Member, IEEE, Ilya Charaev, and Michael Siegel

*Abstract*—The concept of the radio-frequency superconducting nanowire single-photon detector (RF-SNSPD) allows frequency-division multiplexing (FDM) of the bias and readout lines of several SNSPDs. Using this method, a multi-pixel array can be operated by only one feed line. Consequently, the system complexity as well as the heat load is significantly reduced. To allocate many pixels into a small bandwidth the quality factor of each device is crucial.

In this paper, we present an improved RF-SNSPD design. This new design enables a simple tuning of the quality factor as well as the resonant frequency. With a two-pixel device we have demonstrated the operation without crosstalk between the detectors and showed the time, spatial and photon number resolution. Thereby a single pixel requires only a bandwidth of 14 MHz.

## I. INTRODUCTION

Arrays of superconducting nanowire single-photon detectors (SNSPDs) provide fast and effective detection of photons combined with high spectral-, energy and photon-number resolution [1]. These properties make them perfectly suited for optical communication over long distances with high data rates [2], high resolution single-photon imaging [3], spectroscopy applications [4], and many more. However, for all these applications larger pixel numbers are desired which are not easy to implement. The increasing system complexity and heat load leads to an unacceptable level [5], [6]. To handle these challenges several different multiplexing schemes were proposed. By use of RSFQ circuits, current splitting techniques or time-tagged readout methods the number of readout lines can be reduced [7]-[14]. But, there is still no possibility to multiplex the bias lines and apply an individual biasing for each detector in the array.

Several groups successfully demonstrated FDM of large kinetic-inductance detector (KID) arrays [15]-[17]. To adapt a FDM scheme to multi-pixel SNSPD arrays we proposed the usage of the RF-SNSPD [18]. The RF-SNSPD represents a resonant circuit consisting of a lumped-element capacitor C connected to a conventional meander shaped SNSPD as a lumped-element inductor (L). This design is somewhat similar to lumped element KID [19], but the operation principle is different. The resonator operates at 4.2 K and is driven by a microwave current with the amplitude close to the critical current $I_C$ of the nanowire. Thus, the operation regime is highly nonlinear, which enables single-photon detection. In contrast to KIDs, the complete damping of the resonance due to photon-assisted hot-spot formation across the nanowire rather than the change of the kinetic inductance $L_{kin}$ serves as an observable. The jump of the damping inside the circuit can be measured as an increasing transmission coefficient $|S_{21}|$ of the feed line at $f_{res}$. Since this change in transmission is like a binary switching event due to highly nonlinear operation regime, it is much stronger than any noise of the readout electronics.

The RF-SNSPD offers the possibility of a FDM scheme to reduce several bias and readout lines to a single feed line. Using this method, the spatial and temporal information of each detector is provided. However, the available bandwidth of FDM readout systems is limited. Thus the frequency spacing between two RF-SNSPDs needs to be sufficiently small. In order to achieve this goal, increasing quality factors of the RF-SNSPD are required which are not that easy to realize due to the influence of the nonlinear kinetic inductance of the detector. In this paper, we present an optimized microwave design of the RF-SNSPD. Using this new design the quality factor can be increased without compromising the operation of the detector.

## II. DESIGN AND SIMULATION

All structures in this paper have been designed and simulated using Sonnet em [20]. The properties of R-plane sapphire ($\varepsilon_r$ = 10.06 and tan$\delta$ = $10^{-9}$) as substrate with a thickness (d) of 330 µm have been taken into account. As superconducting material we used NbN with d = 5 nm ($R_s$ = 26 µΩ @ f = 5 GHz) which is the standard material of choice for SNSPDs operated at 4.2 K.

The modeling of the RF-SNSPD as a simple series or parallel resonant circuit is not optimal. With increasing bias power $P_B$ the value of the kinetic inductance will exponentially grow. DC-biased NbN nanowires revealed an approximately 6 % increase in the kinetic inductance $L_{kin}$ at a current bias level which corresponds to 80 % of the critical current [21].

This work was supported in part by the Karlsruhe School of Optics and Photonics (KSOP).

S. Doerner, A. Kuzmin, S. Wuensch, I. Charaev, and M. Siegel are with the Institute of Micro- and Nanoelectronic Systems, Karlsruhe Institute of Technology, 76187 Karlsruhe, Germany (steffen.doerner@partner.kit.edu).

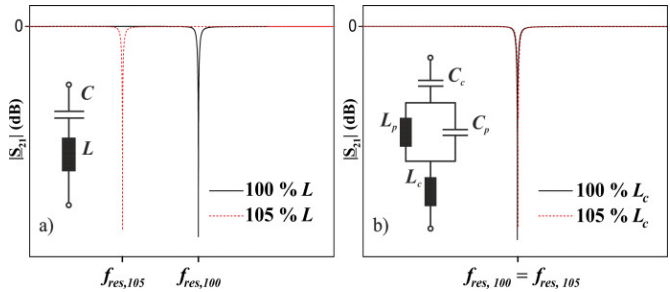

Fig. 1. Simulated influence of the biasing power of an RF-SNSPD onto its resonant frequency. (a) In a series RF-SNSPD configuration with the $L_{kin}$ of the photon sensitive nanowire as the inductive part (L) a change of the inductance about 5 % due to the applied biasing power has a strong influence on $f_{res}$. (b) In the improved RF-SNSPD design with two inductive elements the influence of a changed $L_{kin}$ of the nanowire ($L_c$) can be neglected.

Transferred to RF-SNSPDs such level of biasing would lead to a significantly decreased resonant frequency compared to $f_{res}$ in the zero biased state. If the quality factor is small enough this does not affect the operation of the detector. However, with increasing quality factors the change of $L_{kin}$ leads to an impairment of the detector function depicted in Figure 1a. Without any bias the resonant frequency of the series resonator is observed at $f_{res,100}$. If we take the biasing into account, the stored energy in the resonant circuit will lead to an increasing $L_{kin}$ of the nanowire. For a 5 % increased inductance the resonant frequency will be shifted to $f_{res,105}$. In this state the detector will be photon sensitive. However, after a photon is absorbed and a hotspot is formed the stored energy will be dissipated as in the model of the Duffing nonlinear resonator [22]. As a consequence the resonant frequency will be shifted back to $f_{res,100}$. Further biasing of the detector with a frequency of $f_{res,105}$ will no longer affect the resonant circuit. To bring the detector back into the photon sensitive state the frequency needs to be swept to $f_{res,105}$ again starting from $f_{res,100}$. For the implementation of large arrays this hysteresis effect results in major challenge for the biasing circuits and increases the complexity significantly.

As a possible solution, we developed a new microwave design for the RF-SNSPD shown in Fig. 1b. The whole device is made from a single layer NbN which allows an easy integration into the standard SNSPD fabrication process. The idea is to use a parallel circuit of $L_p$ and $C_p$ to define the resonant frequency. To suppress a changing value of the kinetic inductance $L_p$ due to the applied biasing the line width (w) of $L_p$ needs to be much wider than w of the photon sensitive nanowire. This nanowire is placed as a coupling element ($L_c$) between the parallel circuit and the ground plane. In consequence, the critical current of $L_p$ is much higher than $I_C$ of the photon sensitive nanowire. Because the biasing levels required to operate the detector are defined by $L_c$ the variation of the kinetic inductance of $L_p$ is negligible. However, the kinetic inductance of $L_c$ will still depend on the microwave biasing power. To reduce this influence on the resonant frequency $L_c$ needs to be much smaller than $L_p$. As a second coupling element the capacitor $C_c$ connects the resonant circuit with the feed line and defines mainly the loaded quality factor of the device.

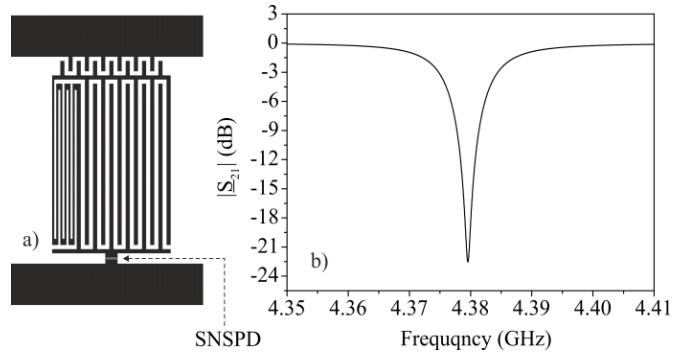

Fig. 2. a) Sketch of the improved RF-SNSPD device embedded in the gap of a coplanar feed line. b) Simulated $|S_{21}|$ coefficients over frequency.

Figure 2a depicts a sketch of the new RF-SNSPD design. The overall circuit with dimensions of 45 x 100 µm is embedded into the gap of a coplanar feed line with characteristic impedance $Z_L$ = 50 Ω. The resonant parts have been simulated using the values $L_p$ = 22 nH, $L_c$ = 7 nH, $C_p$ = 50 fF and $C_c$ = 10 fF. Figure 2b shows the simulated $|S_{21}|$ measurement. The achieved loaded bandwidth measured at -3 dB is almost 9.6 MHz.

III. MEASUREMENT AND RESULTS

To check the simulation and to proof the photon sensitivity of the improved RF-SNSPD design we fabricated several devices of the design shown in Figure 2a. NbN (d = 5 nm) was deposited by reactive magnetron sputtering on R-plane sapphire substrate in Argon/Nitrogen gas atmosphere. The patterning was done with electron-beam lithography and ion-beam etching. The photon sensitive meanders are 16 µm long and 100 nm wide. The active area has a size of 1 x 4 µm with a filling factor of almost 50 %. We note that this design of the nanowire used in our proof-of-principle devices is not optimal in terms of optical coupling and filling factor. However, with keeping the ratio between $L_p$ and $L_c$ the size of the active area can be easily increased in future devices.

For measurements the devices were mounted into a gold-coated brass housing and cooled down to the liquid helium temperature of 4.2 K. A multimode optical fiber was used to guide the photons to the detector. The fiber tip was placed just above the active area. The coaxial cables to the detector block included cold attenuators to reduce the biasing power by -32 dB. On the other side in the return path the detector response was increased again by 39 dB with a cold low-noise amplifier.

A. Single-pixel device

Fig. 3a shows the measured $|S_{21}|$ parameters over frequency of the new RF-SNSPD design. To ensure that we measure the true nonlinear shift of $f_{res}$ we used a vector network analyzer (VNA) and swept the frequency downwards. A measurement with upwards sweeping frequencies is not suitable to measure the increasing kinetic inductance. The resonant frequency will be shifted to lower values while the generator frequency is sweeping to higher values [23]. At very low levels of $P_B$ far away from the critical power $f_{res,0}$ was measured at 4.429 GHz. The loaded bandwidth reached 10.8 MHz which is comparable to the simulation results of Fig. 2b. With increasing $P_B$ only a slight shift of $f_{res}$ can be observed. Just below $P_C$ we measured a shift of 3.2 MHz which is a total change of only 0.072 %. This shift can be further decreased by increasing the ratio between $L_p$ and $L_c$.

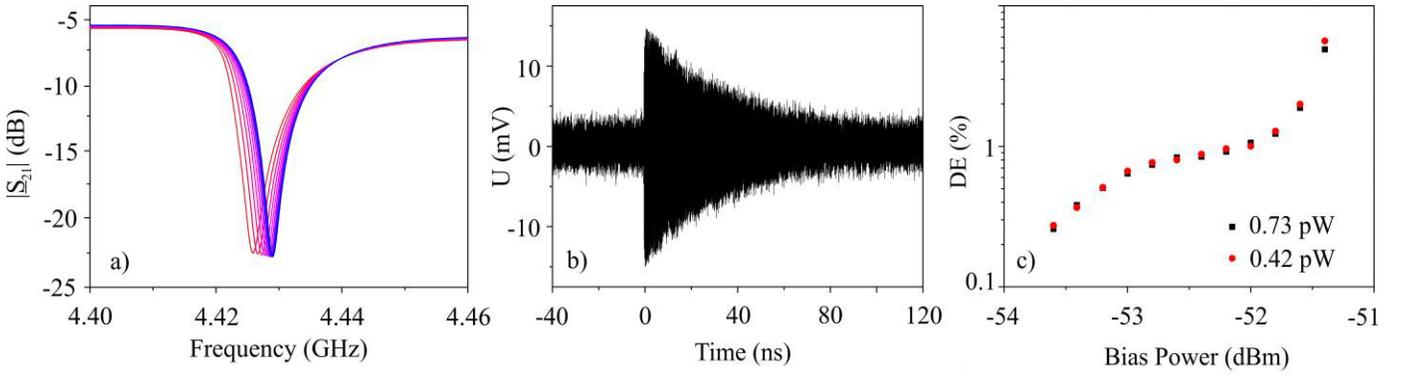

Fig 3. a) Measured $|S_{21}|$ coefficients over frequency of the improved RF-SNSPD device. The change of $f_{res}$ results from the increasing $L_{kin}$ with increasing $P_B$. Just before the critical power is reached the shift of $f_{res}$ to the zero biased state is almost 3.2 MHz. b) Measured voltage over time on the feed line while operating the device at $f_{res}$. For t = 0 ns the $|S_{21}|$ parameter is changed because of an absorbed photon. c) Detection efficiency over biasing power for two different intensities of the incident photons onto the active area. The wavelength of the light was set to 400 nm.

In order to investigate the single-photon response we changed our setup. We replaced the VNA with a microwave signal generator and a real-time oscilloscope. With this experimental setup we are able to record the fast changing $|S_{21}|$ parameter of the feed line at $f_{res}$ during a detection event. In Fig. 3b the single-photon response of the detector is shown. With the forming of a normal conducting domain in the nanowire the quality factor of the resonator is strongly decreased, which can be recognized in the sharp increase of the transmitted signal on the feed line at t = 0 ns. As soon as the superconducting state of the nanowire is recovered the resonant circuit starts with its transient oscillation (t > 0 ns), which is in good comparison to [18]. Hence, the maximum achievable count rate of the device is mainly limited by the transient oscillation time, which is defined by the quality factor of the resonator as $\tau \sim Q/f_{res}$. However, because the quality factor is almost reduced to zero as long the resistive domain persists in the meander, we expect the timing jitter to be independent of the resonator design and in the range of conventional SNSPDs. But, due to our setup we could not measure correct timing jitter values of our devices so far.

In Fig. 3c we also show the measured dependence of the detector's detection efficiency (DE) on biasing power at two different intensities of incident light onto the active area of the device. The DE was calculated as the ratio of measured count rate divided by the number of photons per second, incident onto the nanowire. The wavelength of the photons was set to 400 nm. For low powers an increasing DE with increasing bias power can be observed. With a further increase of $P_B$ the DE starts to saturate before the dark count rate dominates the measured count rates. The achieved results are in good agreement to DC biased SNSPDs. The achieved DE of one percentage can be further increased by optimization of the optical coupling (e.g. AR-coatings and optical cavities). Moreover, the fact that the DE is independent of the number of incident photons demonstrates the single-photon operation regime of the detector.

B. Two-pixel device

To prove the potential of the RF-SNSPD we fabricated a two-pixel device of the design shown in Fig. 2a. Both resonant circuits have the same dimensions. Only $L_p$ is varied in length to tune $f_{res}$. The pixels are located in the gap of the common coplanar feed line. Fig. 4 shows the transmission measurement using the VNA. The resonant frequency of the first pixel $f_{res,1}$ is 4.155 GHz. The resonance of the second pixel is observed at $f_{res,2}$ = 4.327 GHz. With $P_B > P_C$ both resonant events disappear and the $|S_{21}|$ parameter at $f_{res,1}$ and $f_{res,2}$ increases about 18 dB.

To operate both pixels simultaneously we had to use two biasing harmonic signals (tones) at $f_{res,1}$ and $f_{res,2}$ which are continuously generated by two synchronized signal generators. A wide-band signal combiner at room temperature couples both tones into the feed line. In this configuration, the biasing level of each pixel can be individually adjusted. The transmitted signal on the feed line is still measured with the real-time oscilloscope. Because there are now two active pixels the measurement in the time domain is not sufficient. To determine which pixel triggered the event when a microwave pulse on the feed line is observed, a transformation into the frequency domain is required. Therefore, we transmit the data of the oscilloscope continuously to a read-out system. If an event on the feed line is measured the relevant time span is captured. This data is then transformed into the frequency domain using a fast Fourier transformation (FFT) algorithm.

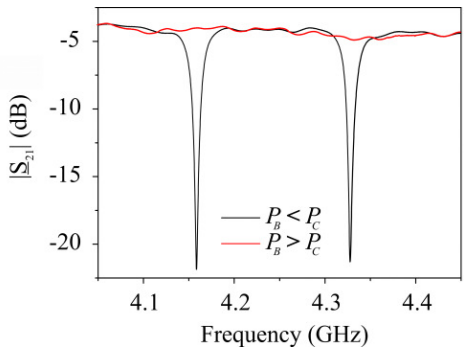

Fig. 4. Measured $|S_{21}|$ coefficients of a two pixel RF-SNSPD device over frequency. The biasing power was adjusted below the critical power and above.

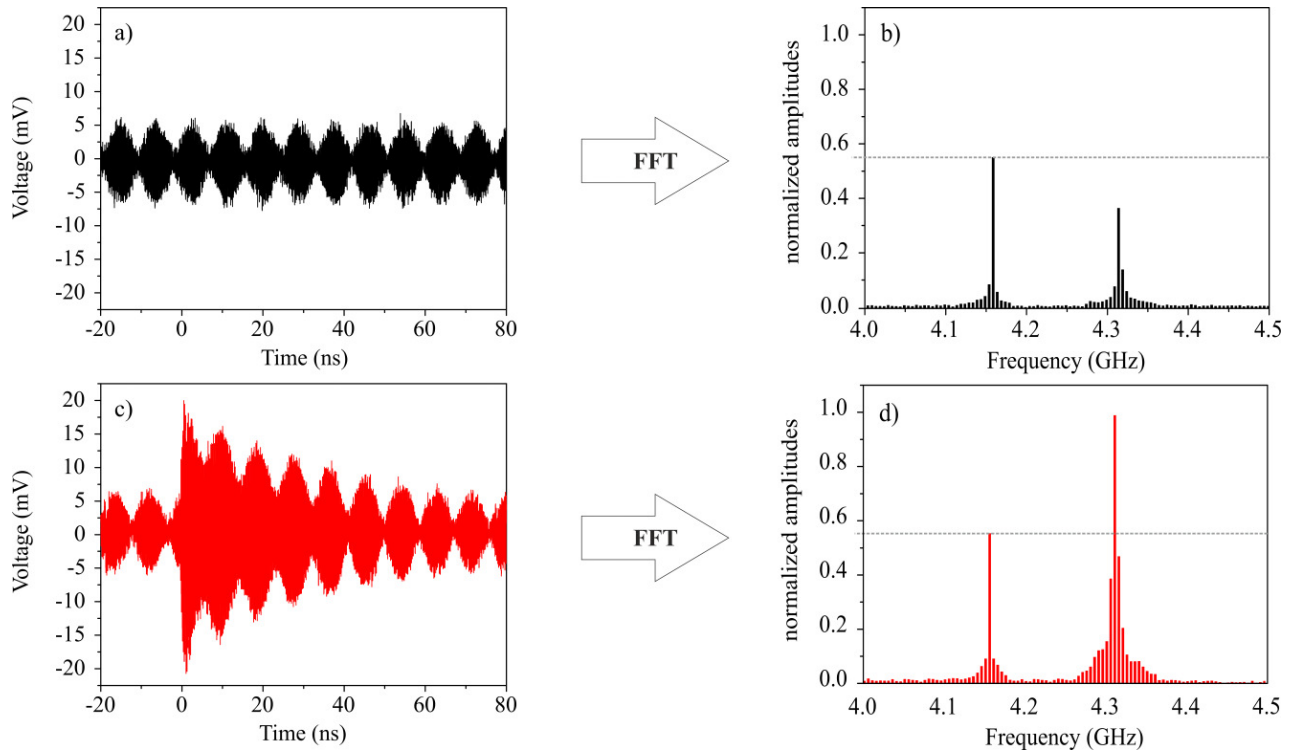

Fig. 5. a,c) Measured voltage over time on the feed line. a) Without an event c) with count event of a single detector. b,d) transformation of the signal a) and c) into the frequency domain. The different signal amplitudes of $f_{res,1}$ and $f_{res,2}$ in b) results from an individual bias level for each pixel. The increased signal at 4.33 GHz in d) reveals the $f_{res}$ of the RF-SNSPD which triggered the event.

In Fig. 5 the readout process is depicted. Fig. 5a shows two biasing tones producing characteristic amplitude modulation. Fig. 5b shows the spectrum of this signal obtained by FFT. Since both pixels had different measured critical bias-power levels we adjusted the power of the tones for each pixel individually to achieve approximately equal DE. Thus, the signal amplitude of $f_{res,2}$ is lower compared to $f_{res,1}$. In Fig. 5c the detector response in case of an absorbed photon in one of the two RF-SNSPDs is shown. Again a beating modulation caused by interference of the two tones can be observed. But, at certain moment also additional transient amplitude modulation appears. In the frequency domain (Fig. 5d) the significant increase of the tone amplitude at $f = f_{res,2}$ is identified. The level of $f_{res,1}$ is the same compared with the measurement without an detection event. It is therefore evident, that the RF-SNSPD operating at $f_{res,2}$ triggered the event. Using this readout method, the exact timing of each event can be measured in the time domain and the information of the spatial and photon-number resolution is given in the frequency domain.

The two-pixel device was also tested for crosstalk during operation of both pixels. In this case we define crosstalk as the effect that one detector counts a false event because of a detection event in the other RF-SNSPD. Since it is not possible with our setup to illuminate the detectors individually, we operated both detectors at a low bias power ($P_B < 0.75\ P_C$). Therefore, the probability of a photon detection event of both pixels at the same time is minimized.

The active areas of the two pixels were illuminated equally with photons of 400 nm wavelength. The gating time using the FFT readout method was set to 10 ns. After more than 15000 detection events only 46 times both detectors triggered an event in the same time window. Thus, the probability of crosstalk is insignificant.

## IV. Conclusion

In summary, a new RF-SNSPD microwave design was developed, simulated and measured. This new design allows the operation of RF-SNSPDs with high quality factors without losing the self-resetting behavior. The shift of the resonant frequency caused by the biasing power is reduced to 0.072 %. Thus, we demonstrated that a single-pixel requires only 14 MHz bandwidth during operation. Hence, more than 70 RF-SNSPDs can be multiplexed within a bandwidth of 1 GHz.

Furthermore, we demonstrated the operation without crosstalk of a two-pixel RF-SNSPD array with a time, spatial and photon-number resolution. With only one single feed line each pixel can be individually biased and readout. The used FDM method provides a simple increase of the pixel numbers in future applications.

## Acknowledgment

The authors would like to thank K. Ilin for helpful discussions during this research, A. Stassen for careful preparation of the samples and also K. Gutbrod for excellent mechanical assistance. This work was supported in part by the Karlsruhe School of Optics and Photonics (KSOP).